# Covariance in protein multiple sequence alignments using groups of columns

Kyle E. Kreth and Anthony A. Fodor
Department of Bioinformatics and Genomics
University of North Carolina at Charlotte, Charlotte, North Carolina 28223, USA.

**ABSTRACT**

Algorithms that detect covariance between pairs of columns in multiple sequence alignments are commonly employed to predict functionally important residues and structural contacts. However, the assumption that covariance only occurs between individual residues in the protein is more driven by computational convenience rather than fundamental protein architecture. Here we develop a novel algorithm that defines a covariance score across two groups of columns where each group represents a stretch of contiguous columns in the alignment. We define a test set that consists of secondary structure elements (α-helixes and β-strands) across more than 1,100 PFAM families. Using these alignments to predict segments that are physically close in structure, we show that our method substantially out-performs approaches that aggregate the results of algorithms that operate on individual column pairs. Our approach demonstrates that considering units of proteins beyond pairs of columns can improve the power and utility of covariance algorithms.

## 1 INTRODUCTION

One of the "grand challenges" (Bradley and Baker, 2006) of structural genomics is to elicit structural information from sequence alone. The relationship between compensatory changes (*e.g.* mutations) of amino acids within structurally constrained regions of homologous proteins has been an active area of research since the pioneering work of Altschuh (Altschuh, et al., 1988). To date, most algorithms use pairs of single columns as the unit of covariation. Covariance between pairs of columns has been used to find errors in alignments (Dickson and Gloor, 2012), locate points of inter-protein docking (Little and Chen, 2009), and to search for packing specific to α-helixes to α-helixes distances (Hopf, et al., 2012). Approaches for these algorithms have varied with scoring based on substitution matrices (Ashkenazy and Kliger, 2010; Gobel, et al., 1994), chi-squared tests (Kass and Horovitz, 2002), perturbation (Dekker, et al., 2004; Lockless and Ranganathan, 1999), and more recently for large multiple sequence alignments (MSA) the inverse of sparse covariance estimations (Jones, et al., 2012). Recent improvements in these algorithms have been substantial (Livesay, et al., 2012), though the basis for improvements have varied significantly including machine learning (Cheng and Baldi, 2007; Jones, et al., 2012), explicit incorporation of information from protein structures (Eyal, et al., 2007), and phylogeny based corrections (Dunn, et al., 2008).

There is no *a priori* reason to think that covariance is limited to individual pairs of residues. A number of researchers therefore have explored methods beyond simple pairs of columns. These methodologies often work with groups of columns (GOC) that are not contiguous within sequence. In one example of this approach, Halabi *et al.* (Halabi, et al., 2009) utilized what they termed "Sectors" in which information from the SCA algorithm is expanded to multiple sets of columns. In another example, Burger *et al.* (Burger and van Nimwegen, 2010) utilized a graph based model that relies on Bayesian statistics to score sets of inter-related columns.

Stretches of residues that are continuous within the protein sequence around structural and functional sites are often conserved (Aloy, et al., 2001; Hoberman, et al., 2004; Hu, et al., 2000; Ouzounis, et al., 1998). It is therefore reasonable to believe that algorithms that work on these continuous GOCs could provide insights into proteins that would be missed by algorithms that work on pairs of columns or on discontinuous sets of columns. With this in mind, we developed an algorithm that detects covariance in these continuous stretches of sequence.

Since the number of permutations for arbitrary non-overlapping GOC for even a modestly sized protein is exponentially large, we developed a test set for our algorithm that focuses on secondary structure elements (SS), specifically α-helixes and β-strands. This approach is attractive because secondary structure elements are predefined, obviously relevant to structure, non-overlapping and modest in number. As demonstrated below, in finding SS elements that are physically close in PDB structures, our approach significantly outperforms methods that aggregate covariance results from pairs of columns.

## 2 METHODS

### 2.1 Covariance algorithms applied to individual pairs of columns and groups of columns.

The goal of this paper is to compare algorithms that calculate covariance on a pair of contiguous "groups of columns" (GOC) within a protein multiple sequence alignment. These algorithms extend algorithms that calculate covariance on a pair of columns. The following algorithms are evaluated in this paper:

**Average McBASC**

The McLachlan (Olmea, et al., 1999) based Substitution Correlation (McBASC) algorithm works on a single pair of columns and has been previous described (Fodor and Aldrich, 2004). Briefly, if N is the number of sequences in the alignment, to calculate a covariance score for columns $i$ and $j$, we create a vector of length $\binom{N}{2}$ for each column. With $k$ and $l$ defined as indexes of each sequence within the alignment, each vector is populated with the values of scores from the McLachlan substitution matrix that result from comparing the residues within each column for all possible comparisons of sequences $k$ and $l$ (with $k \neq l$). The McBASC score $r$, for a given $i,j$ column combination is given by:,

$$r_{i,j} = \frac{1}{N^2} \cdot \frac{\sum_{kl}(s_{ikl}-\langle s_i \rangle)(s_{jkl}-\langle s_j \rangle)}{\sigma_i \sigma_j} \quad (1)$$

where $\langle S \rangle$ is the average and $\sigma$ is the standard deviation for all the entries in each of the two vectors (Fig 1A). $r$ can range from -1 to 1 inclusive with the highest score indicating the highest level of covariance. If either of the columns is perfectly conserved, $r$ is assigned a score of 1. For performance reasons, our implementation (https://github.com/afodor/cobs) produces values of r that are approximate to Eq. 1 (with differences for alignments of >50 sequences of less than 1% from the value of $r$ defined in Eq. 1).

The "McBASC Average" as indicated by the name is the result of taking the average McBASC score for each pair of column within two GOCs. That is, if there are $g_1$ columns in one GOC and $g_2$ columns in another GOC, the "Average McBASC" score is defined as the mean of the $g_1 \cdot g_2$ McBASC scores produced by calculating McBASC for all $g_1$ versus $g_2$ columns.

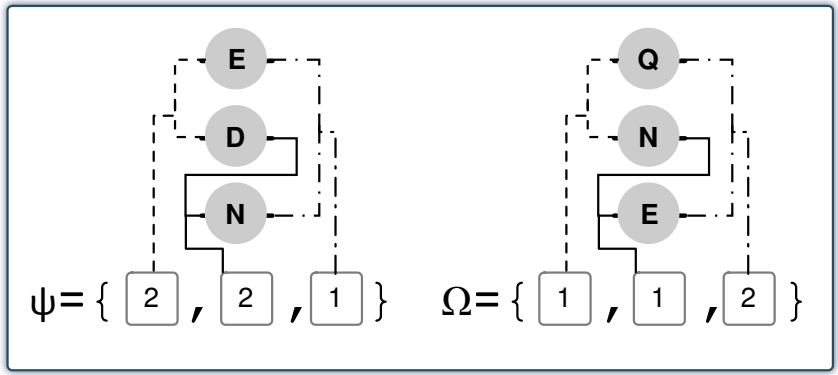

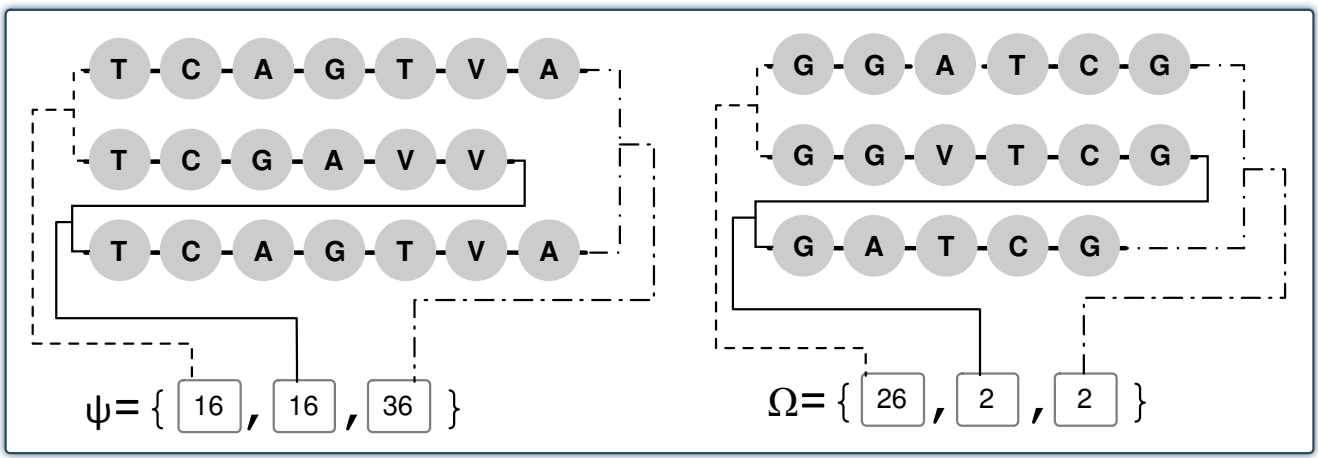

**Fig. 1. McBasc and COBS applied to simple alignments.** (a) The McBasc algorithm applied to two columns from a multiple sequence alignment. The similarity of each pair of amino acids in each column is recorded using a McLachlan matrix. Each score of similarity from the McLachlan matrix is then added to a vector $\psi$ (for the first column) and $\Omega$ (for the second column). (b) The COBS algorithm applied to two contiguous groups of columns within an alignment. The scores added to the vectors for each pair of sequences in the alignment is the sum of all substitutions from the McLachlan matrix.

## **COBS**

As a simple alternative to averaging all possible McBASC scores within two GOCs, we propose COBS (COvariance By Sections), a straight-forward extension of McBASC to groups of contiguous columns (Fig. 1). As in McBASC, we end up with a pair of vectors which are compared by Pearson correlation to give a final score. If $i$ represents a GOCs of length $m$ within the alignment, and $k$ and $l$ are indexes of each sequence within the alignment, then the value placed within the vector for $i$ is given by:

$$S_{ikl} = \sum_m \text{McLachlan}(k_m, l_m) \quad (2)$$

where the McLachlan function returns the substitution matrix value comparing the residues at position *m* within the GOC for sequences *k* and *l* (Fig. 1B). To generate a COBS score for GOC *i* vs. GOC *j*, vectors of length $\binom{N}{2}$ are generated for each GOC and over all possible comparisons of *k* and *l* (with *k* != *l*), the two vectors are populated with the values generated by Eq. 2. The vectors are scored by the Pearson correlation as indicated by Eq. 1 to generate a final COBS score. GOCs that are perfectly conserved are given a score of 1.

## Average Conservation

We calculated Shannon Entropy as canonically defined (Shenkin, et al., 1991):

$$-\sum_{x_1}^{x_{20}}(p_x(i)\ln p_x(i)) \quad (3)$$

with *x* being indexed across all 20 amino acids, $p_x$ representing the frequency of the particular amino acid at the $i^{th}$ column. The "Conservation Average" is the mean value for this value across all the columns in the pair of GOCs.

## Mutual Information

Mutual information was implemented as previously described (Fodor and Aldrich, 2004). As was the case for average McBASC, we define average MI as the mean of the $g_1 \cdot g_2$ MI scores from 2 groups of columns (with $g_1$ columns in the first group and $g_2$ columns in the second group).

**2.2 Phylogenetic Correction**

MI has been to shown to be an ineffective measure of covariance within protein alignments (Fodor and Aldrich, 2004) with a high sensitivity to phylogenetic artifacts in the alignment. A procedure to correct for these artifacts has been introduced (Dunn, et al., 2008) and been shown to substantially improve the performance of MI. If MI scores have, as indicated above, been calculated for all pairs of columns *i* and *j* in the alignment, then MI with a phylogenetic correction (Dunn, et al., 2008) termed MIp is calculated as:

$$MIp(i,j) = MI(i,j) - \frac{MI(i,\bar{x}) \cdot MI(j,\bar{x})}{\overline{MI}} \quad (4)$$

where $MI(i,\bar{x})$ is the average MI score of column *i* with all other columns in the alignment, $MI(j,\bar{x})$ is the average MI score of column *j* with all other columns in the alignment and $\overline{MI}$ is the average of all MI scores from all pairs of columns in the alignment.

The APC correction has been shown to work with McBASC, at least for small alignments (Buslje, et al., 2009) previously. We use the same correction on McBASC as we defined for MI:

$$McBASCp(i,j) = McBASCp(i,j) - \frac{McBASC(i,\bar{x}) \cdot McBASC(j,\bar{x})}{\overline{McBASC}} \quad (5)$$

where $McBASC(i, \bar{x})$ is the average McBASC score of column $i$ with all other columns in the alignment, $McBASC(j, \bar{x})$ is the average McBASC score of column $j$ with all other columns in the alignment and $\overline{McBASC}$ is the average of all MI scores from all pairs of columns in the alignment.

As with "Average McBASC" and "Average MI", we define "Average MIp" and "Average McBASCp" as the mean of the $g_1 \cdot g_2$ MIp or McBASCp scores respectively from 2 groups of columns (with $g_1$ columns in the first group and $g_2$ columns in the second group).

Although implemented originally for use with algorithms that work on pairs of columns, the phylogenetic correction algorithm can be applied to any other covariance score (Burger and van Nimwegen, 2010; Savojardo, et al., 2013). If we have Y GOCs in our dataset, we will generate a total of $\binom{y}{2}$ COBS scores. For each pair of GOCs, $i$ and $j$, the phylogenetic corrected COBS score (which we call COBSp) is given by:

$$COBSp(SS_i, SS_j) = COBS(SS_i, SS_j) - \frac{COBS(SS_i\bar{x}) \cdot COBS(SS_j, \bar{x})}{\overline{COBS}} \quad (6)$$

where $COBS(i, \bar{x})$ is the average COBSs score between GOC $i$ and all other GOCs in the alignment, $COBS(j, \bar{x})$ is the average COBS score for GOC $j$ and all other GOCs in the alignment and $\overline{COBS}$ represents the average of all COBS scores for the alignment in question.

As an additional permutation, the APC correction for MI and McBASC can be applied on a per column basis as defined above and was originally done for Mutual Information, or can be done on GOCs, which is what COBS must use since it only works at the GOC level. In our implementation (https://github.com/afodor/cobs), we output phylogenetic correction for McBASC and MI at the pair of column level (before pairs of columns are averaged), at the group of column level (after pairs of columns are averaged) and applied twice: initially at the pair of columns and then again at the group of columns level. None of these normalization schemes consistently out-perform any of the others for McBASC and MI (data not shown). In this paper, therefore, we only report normalization at the single pair of column level (before pairs of columns are averaged) which is most consistent with how phyolgenetic corrections have been previously utilized in the literature.

## 2.2 Source Data and Distance Computation

Version 26 (November 2011) of PFAM (Boeckmann, et al., 2003) was downloaded from ftp.sanger.ac.uk. Protein families were chosen that had at least one protein referenced in the GF DR line. For sampling size considerations, only families that had at least 200 sequences were considered. For performance considerations families with > 2000 sequences were removed from the data set. To ensure that we would have enough columns for analysis, MSA "width" was set to a floor of 80 columns. PDB structures were assigned to PFAM families based on Sanger mappings (ftp://ftp.sanger.ac.uk/pub/databases/Pfam/mappings/pdb_pfam_mapping.txt). SS elements were determined using the "HELIX" or "SHEET" indicators in the remarks section of the selected PDB files, which were downloaded from rcsb.org. Distances between all β-carbons for a given SS were measured against all the β-carbons of the other SS being compared. In the cases where Glycine was part of the measurement α-carbons were used.

Mapping of each PDB file to each PFAM alignment was achieved using a BLAST search of the PDB sequence against all sequences in the PFAM alignment taking the best hit as our reference sequence. A Needleman–Wunsch alignment was then performed between the PDB sequence and the PFAM reference sequence and this alignment was used to map PFAM columns to the corresponding structural residues. In total 1,116 PFAM families were found that had matching PDB files that had a minimum of 7 secondary structure elements and a percent identity between the PDB and PFAM reference sequence of at least 90% in the Needleman–Wunsch global alignment. The protein families that were used in this study are available at https://github.com/afodor/cobs/blob/master/cobs/pdbToPfamForCobsViaBlast.txt.gz.

In generating ROC curves, we simply took the scores for each PFAM family and aggregated them into one large spreadsheet sorted by score. This method has the disadvantage that the top hits that represent the initial set of predictions from the ROC curve may come from a disproportionally small number of PFAM families. An alternative would be to generate a separate ROC curve for each PFAM family and then produce an average ROC curve made up of each individual ROC curve. However, this procedure generated nearly identical ROC curves as simply taking absolute score (data not shown). In figures for this paper, therefore, we report ROCs based on absolute scores. Results for the alternative method can be generated by following the final step in the "Readme" instructions for the code for this paper available at https://github.com/afodor/cobs/.

## **RESULTS**

### 2.3 Considering groups of columns explicitly yields improved predictions of physically close secondary structure elements

We defined a novel covariance method called COBS that works on contiguous groups of columns within a protein multiple sequence alignment (see methods). We evaluated the COBS algorithm on the proximity of secondary structure (α-helixes and β-strands) in 1,116 PFAM families. For each PFAM alignment, we asked in the corresponding structure how well the COBS algorithm could predict secondary structure elements that were in physical proximity. We compared the COBS algorithm to averaging results from the canonical covariance methods MI and McBASC that work on individual pairs of columns (see methods).

Fig. 2 shows the results of this comparison for the Bac_rhamnosid PFAM family (*PF05592*). There are 18 α-helixes and 17 β-strands in this family for a total for 35 secondary structure elements. For each of the 595 (or $\binom{35}{2}$) possible comparisons, we asked how well the scores from the variance covariance algorithms predict the average distance between all residues in these structures. As controls, we included the average conservation score for both columns as well as simply assigning a score from a random (uniform) distribution. Just by visual inspection, for this protein family the highest scoring COBS pair of GOCs (to the right on the x-axis) appear to have an average distance that is closer (to the bottom of the y-axis) then the highest scoring pair of GOCs chosen by average McBASC and average MI.

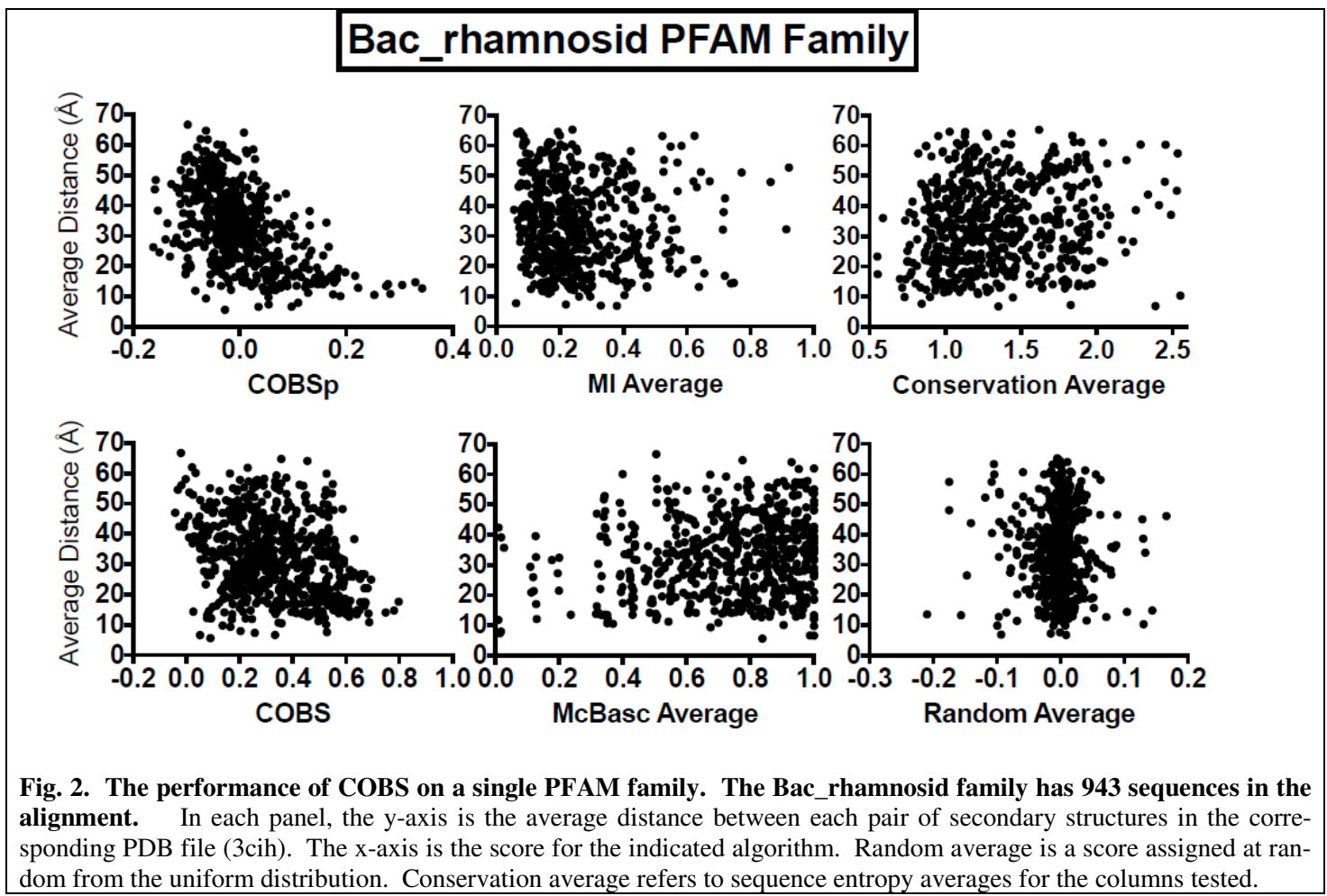

**Fig. 2. The performance of COBS on a single PFAM family. The Bac_rhamnosid family has 943 sequences in the alignment.** In each panel, the y-axis is the average distance between each pair of secondary structures in the corresponding PDB file (3cih). The x-axis is the score for the indicated algorithm. Random average is a score assigned at random from the uniform distribution. Conservation average refers to sequence entropy averages for the columns tested.

In order to gauge the performance of the algorithms across multiple PFAM families, we aggregated all predictions across 180,851 α-helices and β-strands combinations from the 1,116 PFAM families that met the criteria for inclusion in our study (see methods). We arbitrarily defined a success as a prediction in which the average distance between two secondary structure elements is less than the median distance of all secondary structure elements within the protein structure. We then ranked the predictions with the highest scoring prediction first. ROC curves based on these ranks are shown in Fig. 3A. As expected, an algorithm that chooses pairs of secondary structures at random falls on the identity line on the ROC curve (Fig. 3A black line). Average conservation (Fig. 3A, blue line) does little better than random, demonstrating that, unlike for pairs of columns (Fodor and Aldrich, 2004), background conservation does not predict physically close secondary structures. Average MI (Fig. 3A, green line) and average McBASC (Fig 3A, yellow line) are also not much better than random but COBS (Fig. 3A, red line) displays a substantially improved performance.

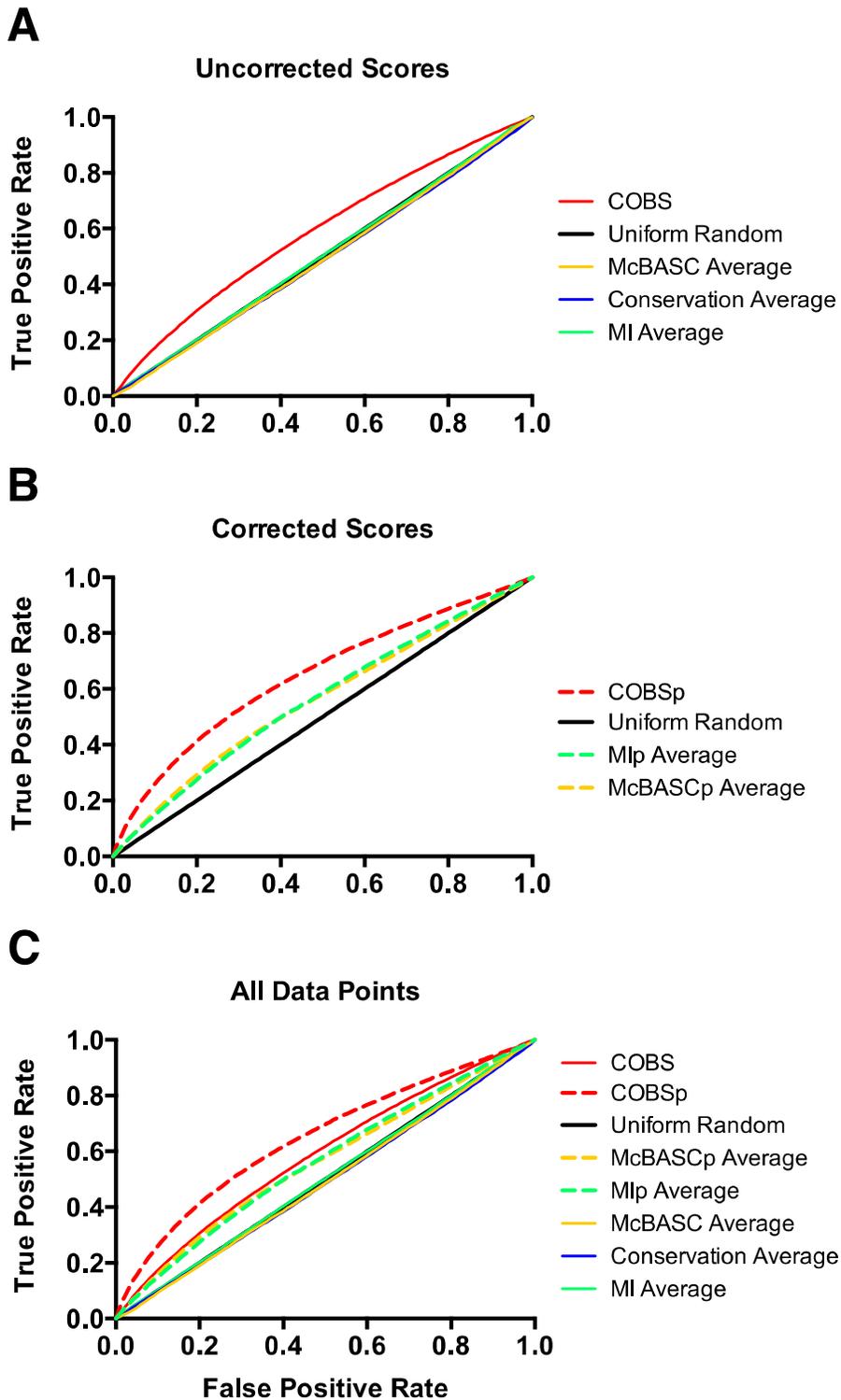

**Fig. 3.** Receiver Operating Characteristic (ROC) curves showing the relative performance for all algorithms. A true positive was defined as any distance that was less than the 50[th] percentile of the average distances of the secondary structures from each alignment. (3A): Algorithms uncorrected for phylogenetic artifacts; (3B): Algorithms with phylogenetic correction applied; (3C): Superimposed ROC curves from corrected (dashed lines) and uncorrected (solid lines) algorithms.

## 2.4 Improved prediction accuracy using Phylogeny correction.

We applied the phylogenetic correction term APC, introduced by Dunn (Dunn, et al., 2008) to the MI, McBASC, and COBS algorithms to produce algorithms called MIp, McBASCp, and COBSp (Fig 3B; see methods). The phylogenetic correction yielded a significant improvement in the performance of MIp (Fig 3B, green dashed line), McBASCp (Fig 3B, yellow dashed line) and COBSp (Fig. 3B, red dashed line) in predicting physically close secondary structures. When all corrected and uncorrected algorithms are compared, COBSp (Fig. 3C, red dashed line) clearly demonstrated the best performance among all the algorithms we tested.

The phylogenetic correction term is designed to eliminate "background" covariance due to non-random sampling across phylogenetic space in the multiple sequence alignment. Since we expect that most secondary structure elements within a protein will not covary, we would expect that after phylogenetic correction, the average covariance score for COBSp would be centered on zero, which would result from a Pearson correlation of unrelated vectors. Fig. 4 demonstrates that this expectation was realized for COBSp score providing further evidence that the simple phylogenetic correction terms is effective in reducing covariance introduced by phyologenetic artifacts.

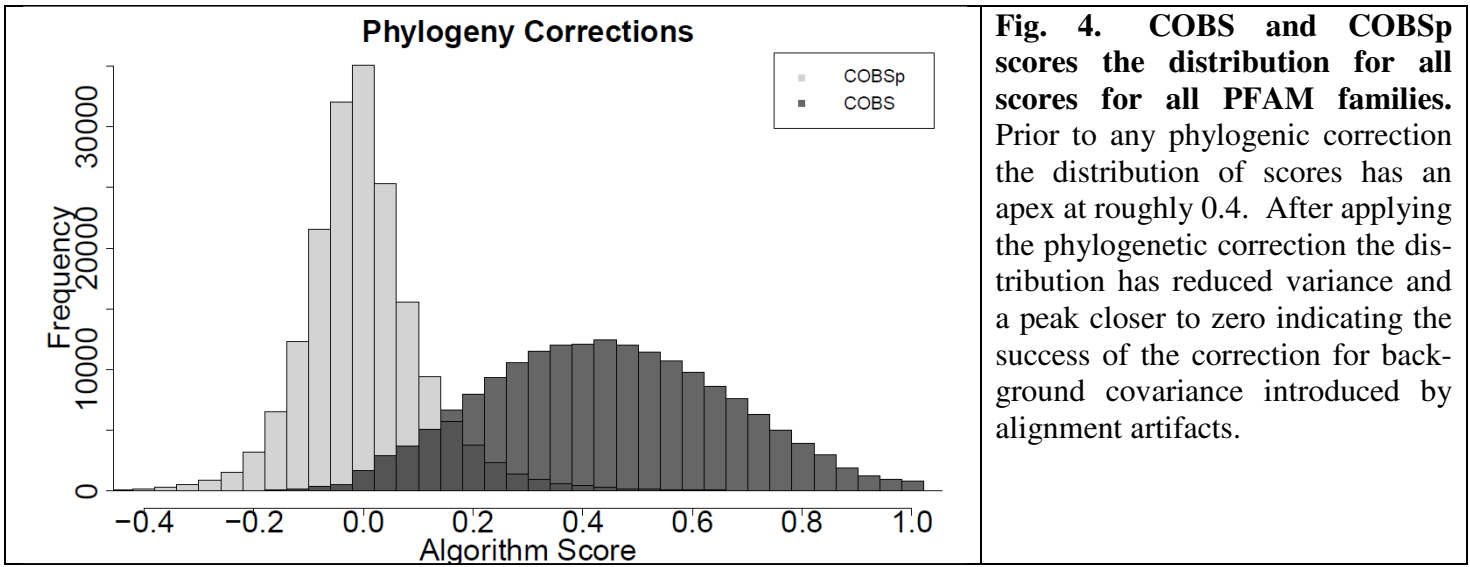

Fig. 4. COBS and COBSp scores the distribution for all scores for all PFAM families. Prior to any phylogenic correction the distribution of scores has an apex at roughly 0.4. After applying the phylogenetic correction the distribution has reduced variance and a peak closer to zero indicating the success of the correction for background covariance introduced by alignment artifacts.

## 3 DISCUSSION

Using an algorithm based in part on average MI scores, Xu and Tillier (Xu and Tillier, 2010) found that residues close to highly covarying residues also tended to be highly covarying. In their work, Xu and Tiller suggest a scoring scheme for a group of residues (what they term a "patch" and what we here call a GOC) based on the MI score for the pair of residues within the patch with the highest covariance score (what they term the "focal pair") divided by the average MI for the entire patch of continuous residues. Here, we suggest an alternative that computes covariance directly at the "patch" or GOC level without relying on average paired covariance.

On a test set of α-helixes and β-strands derived from the PFAM database, our approach appears to have more power at detecting physically close sets of residues than methods that average over covariance scores derived from pairs of columns.

The dataset we used to test our algorithm focused solely on secondary structure covariance. It is easy to imagine future permutations that would extend COBS past α-helixes and β-strands. For example, a "greedy" algorithm could start with "focal pairs" of highly covarying columns and attempt to extend the region of significant covariance. Likewise, since Eq. 2 can be defined over any set of contiguous or non-contiguous columns, one can also imagine possible extensions that could apply COBS to non-contiguous columns to attempt to find a global network of covariance within each protein family. Such extensions, however, would require additional parameters to determine appropriate threshold cutoffs for when groups of covarying columns should be considered distinct clusters. Fitting these additional parameters would presumably require separating part of our data into a training set to estimate the parameters and a separate test set to evaluate performance. By pre-defining our GOCs as secondary structure elements whose composition is defined independent of any action of the algorithms, we have avoided the need for training and test sets, simplifying the interpretation of the relative power of the different algorithms that we tested.

While still modest in overall accuracy, our approach would appear to reveal regional patterns of covariance that are relatively unexplored by algorithms that focus on pairs of columns. This approach may in the future have utility in assisting computational methods that discriminate likely and unlikely folds as well as methods that use sequence alignments to find functionally and structurally important regions in proteins (Aguilar, et al., 2012; Jones, et al., 2012; Othersen, et al., 2012).

**ACKNOWLEDGEMENTS**
We thank Greg Gloor and Dennis Livesay for helpful comments on a previous version of this manuscript.

**References**

Aguilar, D., Oliva, B. and Marino Buslje, C. (2012) Mapping the mutual information network of enzymatic families in the protein structure to unveil functional features, *PLoS One*, **7**, e41430.
Aloy, P.*, et al.* (2001) Automated structure-based prediction of functional sites in proteins: applications to assessing the validity of inheriting protein function from homology in genome annotation and to protein docking, *J Mol Biol*, **311**, 395-408.
Altschuh, D.*, et al.* (1988) Coordinated amino acid changes in homologous protein families, *Protein Eng*, **2**, 193-199.
Ashkenazy, H. and Kliger, Y. (2010) Reducing phylogenetic bias in correlated mutation analysis, *Protein Eng Des Sel*, **23**, 321-326.
Boeckmann, B.*, et al.* (2003) The SWISS-PROT protein knowledgebase and its supplement TrEMBL in 2003, *Nucleic acids research*, **31**, 365-370.
Bradley, P. and Baker, D. (2006) Improved beta-protein structure prediction by multilevel optimization of nonlocal strand pairings and local backbone conformation, *Proteins*, **65**, 922-929.
Burger, L. and van Nimwegen, E. (2010) Disentangling direct from indirect co-evolution of residues in protein alignments, *PLoS Comput Biol*, **6**, e1000633.
Buslje, C.M.*, et al.* (2009) Correction for phylogeny, small number of observations and data redundancy improves the identification of coevolving amino acid pairs using mutual information, *Bioinformatics*, **25**, 1125-1131.
Cheng, J. and Baldi, P. (2007) Improved residue contact prediction using support vector machines and a large feature set, *BMC bioinformatics*, **8**, 113.

Dekker, J.P., *et al.* (2004) A perturbation-based method for calculating explicit likelihood of evolutionary covariance in multiple sequence alignments, *Bioinformatics*, **20**, 1565-1572.
Dickson, R.J. and Gloor, G.B. (2012) Protein sequence alignment analysis by local covariation: coevolution statistics detect benchmark alignment errors, *PLoS One*, **7**, e37645.
Dunn, S.D., Wahl, L.M. and Gloor, G.B. (2008) Mutual information without the influence of phylogeny or entropy dramatically improves residue contact prediction, *Bioinformatics*, **24**, 333-340.
Eyal, E., *et al.* (2007) A pair-to-pair amino acids substitution matrix and its applications for protein structure prediction, *Proteins*, **67**, 142-153.
Fodor, A.A. and Aldrich, R.W. (2004) Influence of conservation on calculations of amino acid covariance in multiple sequence alignments, *Proteins*, **56**, 211-221.
Gobel, U., *et al.* (1994) Correlated mutations and residue contacts in proteins, *Proteins*, **18**, 309-317.
Halabi, N., *et al.* (2009) Protein sectors: evolutionary units of three-dimensional structure, *Cell*, **138**, 774-786.
Hoberman, R., Klein-Seetharaman, J. and Rosenfeld, R. (2004) Inferring property selection pressure from positional residue conservation, *Applied bioinformatics*, **3**, 167-179.
Hopf, T.A., *et al.* (2012) Three-dimensional structures of membrane proteins from genomic sequencing, *Cell*, **149**, 1607-1621.
Hu, Z., *et al.* (2000) Conservation of polar residues as hot spots at protein interfaces, *Proteins*, **39**, 331-342.
Jones, D.T., *et al.* (2012) PSICOV: precise structural contact prediction using sparse inverse covariance estimation on large multiple sequence alignments, *Bioinformatics*, **28**, 184-190.
Kass, I. and Horovitz, A. (2002) Mapping pathways of allosteric communication in GroEL by analysis of correlated mutations, *Proteins*, **48**, 611-617.
Little, D.Y. and Chen, L. (2009) Identification of coevolving residues and coevolution potentials emphasizing structure, bond formation and catalytic coordination in protein evolution, *PLoS One*, **4**, e4762.
Livesay, D.R., Kreth, K.E. and Fodor, A.A. (2012) A critical evaluation of correlated mutation algorithms and coevolution within allosteric mechanisms, *Methods Mol Biol*, **796**, 385-398.
Lockless, S.W. and Ranganathan, R. (1999) Evolutionarily conserved pathways of energetic connectivity in protein families, *Science*, **286**, 295-299.
Olmea, O., Rost, B. and Valencia, A. (1999) Effective use of sequence correlation and conservation in fold recognition, *J Mol Biol*, **293**, 1221-1239.
Othersen, O.G., *et al.* (2012) Application of information theory to feature selection in protein docking, *Journal of molecular modeling*, **18**, 1285-1297.
Ouzounis, C., *et al.* (1998) Are binding residues conserved?, *Pac Symp Biocomput*, 401-412.
Savojardo, C., *et al.* (2013) Prediction of disulfide connectivity in proteins with machine-learning methods and correlated mutations, *BMC bioinformatics*, **14 Suppl 1**, S10.
Shenkin, P.S., Erman, B. and Mastrandrea, L.D. (1991) Information-theoretical entropy as a measure of sequence variability, *Proteins*, **11**, 297-313.
Xu, Y. and Tillier, E.R. (2010) Regional covariation and its application for predicting protein contact patches, *Proteins*, **78**, 548-558.